\documentclass[a4paper,11pt,english]{article}
\usepackage[cp1251]{inputenc}
\usepackage[T2A]{fontenc}
\usepackage[english]{babel}
\usepackage[dvips]{graphicx}
\usepackage{amsmath}
\usepackage{amssymb}
\usepackage{amsxtra}
\usepackage{amsthm,amscd}
\usepackage{hyperref}

\usepackage{geometry}
\newcommand{\mbs}[1]{ {\boldsymbol #1} }
\newcommand {\bbR}{\mathbb{R}}
\newcommand {\bbI}{\bbR^3(h,g,k)}
\newcommand {\rang}{\mathop{\rm rank}\nolimits}

\newcommand{\cons}{\mathop{\rm const}\nolimits}

\numberwithin{equation}{section}
\newcommand{\mysec}[1]{\vspace{5mm}\addtocounter{section}{1}{\bf #1.}\setcounter{equation}{0}}

\begin{document}

\title{Bifurcation of common levels of first integrals\\ of the Kovalevskaya problem}

\author{M.P.\,Kharlamov}

\date{}

\maketitle

\begin{center}
{\bf \textit{Journal of Applied Mathematics and Mechanics}, V. 47, No. 6, 1983, pp. 737--743}
\end{center}

\begin{abstract}
The structure of integral manifolds in the Kovalevskaya problem of the motion of a heavy rigid body about a fixed point is considered. An analytic description of a bifurcation set is obtained, and bifurcation diagrams are constructed. The number of two-dimensional tori is indicated for each connected component of the supplement to the bifurcation set in the space of the first integrals constants. The main topological bifurcations of the regular tori are described.
\end{abstract}

The solution of the problem of the motion of a rigid body about a fixed point found by S.V.\,Kovalevskaya  \cite{bib01} has been considered in many publication. Here we mention only a few of them. G.G.\,Appelrot was the first to point out four special classes of the Kovalevskaya gyroscope motion \cite{bib02}. A more detailed study of these particular motions was fulfilled in \cite{bib03}, where a geometric treatment of Appelrot classes was presented as corresponding to parts of the surface of multiple roots of the Kovalevskaya polynomial in the space of the constants of the first integrals. The hodographs method was used in \cite{bib04, bib05} for a complete study of the motions belonging to the first and second classes, and of the so-called especially remarkable motions of the third class in which the moving hodograph of the angular velocity of the body is a closed curve.

The Appelrot classes correspond to a set of zero measure in the space of the integral constants. All other motions have not been studied to any reasonable extent. Recently in \cite{bib06}, some essential qualitative properties of them were established supposing independence of the first integrals of the Euler\,--\,Poisson equations. Nevertheless, until now it is not known where these integrals are independent. Here we prove that the cases of dependence correspond exactly to the Appelrot classes. Moreover, the study of this question allows us to find out, for all cases, the number of the connected components of integral manifolds; each of the components in the space of the Euler\,--\,Poisson variables is a two-dimensional torus with quasi-periodic motions
\cite{bib07, bib08}. Previously, in \cite{bib06}, it was proved that those manifolds which do not degenerate as the Poincar\'{e} parameter tends to zero consist of two tori.

The investigation of integral manifolds is a part of the solution of the problem of the topological analysis of classical dynamical systems. This problem is due to Poincar\'{e} and Birkhoff. In modern terms, it was formulated by Smale \cite{bib09}.

Let us note also the work \cite{bib10}, where the particular problem of finding the bifurcation set of the energy and momentum integrals is solved. Some inaccuracies of \cite{bib10} were corrected in \cite{bib11} when investigating the general cases. In \cite{bib10}, the Kovalevskaya integral and, consequently, the complete integrability of the system were not considered.

\mysec{1} Let $p$, $q$, $r$ be the components in the trihedral rotating with the rigid body of the angular velocity vector $\mbs{\omega}$, and $\nu_1$, $\nu_2$, $\nu_3$ the components of the unit vertical vector $\mbs{\nu}$. By a suitable selection of the moving axes and units of measurement we reduce the Euler -- Poison equations in the Kovalevskaya problem to the form
\begin{equation}\label{eq1_1}
\begin{array}{c}
    2\dot{p}=qr, \quad 2\dot{q}=-(pr+\nu_3), \quad \dot{r}=\nu_2, \\
    \dot{\nu}_1=r\nu_2-q\nu_3, \quad \dot{\nu}_2=p\nu_3-r\nu_1, \quad \dot{\nu}_3=q\nu_1-p\nu_2.
\end{array}
\end{equation}
The first integrals are
\begin{eqnarray}
& &     2(p^2+q^2)+r^2-2\nu_1=2h, \label{eq1_2}\\
& &    2(p\nu_1+q\nu_2)+r\nu_3=2l, \label{eq1_3}\\
& &    \nu_1^2+\nu_2^2+\nu_3^2=1, \label{eq1_4} \\
& &    (p^2-q^2+\nu_1)^2+(2pq+\nu_2)^2=k. \label{eq1_5}
\end{eqnarray}

Let us recall the main points of Appelrot's classification. We introduce, as in \cite{bib01}, the variables $s_1$, $s_2$:
\begin{equation}\label{eq1_6}
    s_{1,2}=h+\left.\left[R(x_1,x_2)\mp\sqrt{R(x_1)R(x_2)}\right]\right/(x_1-x_2)^2,
\end{equation}
where
\begin{eqnarray}
& & x_1=p+iq, x_2=p-iq, \label{eq1_7}\\
& &     R(x_1,x_2)=-x_1^2x_2^2+2hx_1x_2+2l(x_1+x_2)+1-k, \label{eq1_8}\\
& &     R(x)=-x^4+2hx^2+4l x+1-k. \label{eq1_9}
\end{eqnarray}
The dependence of the variables \eqref{eq1_6} on time is determined by the equations
\begin{equation*}
\begin{array}{c}
    (s_1-s_2)^2 \dot{s}_1^2=-2\Phi(s_1), \qquad (s_1-s_2)^2\dot{s}_2^2=-2\Phi(s_2), \\[2mm]
    \Phi(s)=(s-h+\sqrt{k}(s-h-\sqrt{k})\varphi(s), \quad \varphi(s)=s^2-2hs^2+(h^2+1-k)s=2l^2,
\end{array}
\end{equation*}
where $\varphi(s)$ is the Euler resolvent of the polynomial \eqref{eq1_9}.

Consider the constants of the first integrals for which the polynomial $\Phi(s)$ has a multiple root; Appelrot called the corresponding motions \textit{the simplest motions}. The surface of multiple roots in $\bbI$ consists of the plane
\begin{equation}\label{eq1_10}
    k=0
\end{equation}
(the first class of the simplest motions or the Delon\'{e} class), of the surface
\begin{equation}\label{eq1_11}
    k=(h-2l^2)^2
\end{equation}
(the second and third classes of the simplest motions), and of the surface of multiple roots of the resolvent $\varphi(s)$ described by the equation \cite{bib03}
\begin{equation}\label{eq1_12}
    (1-k)(h^2+1-k)^2-2[9h(1-k)+h^3]l^2+27l^4=0
\end{equation}
(the forth class of the simplest motions). The surface \eqref{eq1_12} may be represented in the parametric form
\begin{equation}\label{eq1_13}
    h=s+\frac{l^2}{s^2},k=1-\frac{2l^2}{s}+\frac{l^4}{s^4}
\end{equation}
or
\begin{equation}\label{eq1_14}
    h=\left.(x^3-l )\right/x, k=1+2l x+x^4.
\end{equation}
In \eqref{eq1_13} the parameter $s$ stands for a multiple root of $\varphi(s)$, and in \eqref{eq1_14} $x$ is a multiple root of the original polynomial \eqref{eq1_9}.

\mysec{2} Let us return to the system \eqref{eq1_2} -- \eqref{eq1_5}. For fixed $k$, $h$, $l$ it defines in $\bbR^6(\mbs{\omega},\mbs{\nu})$ the surface $J_{k,h,l}$, which is invariant under the phase flow \eqref{eq1_1}. This surface is called an integral manifold. Let us recall some known  facts formulated in terms of this problem.

The point $(\mbs{\omega},\mbs{\nu})\in\bbR^6$ is called critical if at this point the rank of the Jacobi matrix is less than four, i.e.,
\begin{equation}\label{eq2_1}
    \rang\left\|
    \begin{array}{cccccc}
        2p & 2q & -1 & 0 & r & 0\\
        2\nu_1 & 2\nu_2 & 2p & 2q & \nu_3 & r\\
        0 & 0 & \nu_1 & \nu_2 & 0 & \nu_3\\
        2(p\eta_1+q\eta_2) & 2(p\eta_2-q\eta_1) & \eta_1 & \eta_2 & 0 & 0
    \end{array}
    \right\|<4.
\end{equation}
Here
\begin{equation}\label{eq2_2}
    \eta_1=p^2-q^2+\nu_1, \qquad \eta_2=2pq+\nu_2.
\end{equation}

A point $(k,h,l)\in\bbR^3$ is called a regular value if on the corresponding surface $J_{k,h,l}$ there are no critical points (in particular, if $J_{k,h,l}=\varnothing$). A point $(k,h,l)$ which is not a regular value is, by definition, a critical value. Critical values fill the set $\Sigma$ in $\bbR^3$ called the bifurcation set.

According to \eqref{eq1_2}, \eqref{eq1_4}, the sets $J_{k,h,l}$  are compact. Therefore, they change the differentiable type only when passing through $\Sigma$. By the Liouville -- Arnold theorem, for $(k,h,l)\in\bbR^3\setminus \Sigma$ the set $J_{k,h,l}$ is the union of a finite number of two-dimensional tori. For the complete investigation of regular integral manifolds it is therefore sufficient to construct the set $\Sigma$ and establish the number of tori in each of the connected components $\bbR^3\setminus\Sigma$.

Let us consider some special cases. In \eqref{eq1_2} -- \eqref{eq1_5} put
\begin{equation}\label{eq2_3}
    q=0.
\end{equation}
Substitute the values of $h$,$k$,$l$ from \eqref{eq1_2}, \eqref{eq1_3}, \eqref{eq1_5} into the polynomial $R(x)$ and its derivative simultaneously assuming $x=p$. We obtain
\begin{equation}\label{eq2_4}
\begin{array}{c}
    R(p)=(pr+\nu_3)^2,\qquad
    R'(p)=2r(pr+\nu_3).
\end{array}
\end{equation}
In the matrix \eqref{eq2_1}, let us change the last row for the combination of rows with the coefficients $p^2,p,1,-1$ respectively. After some elementary transformations (eliminating common factors and transposing rows and columns), we obtain the condition
\begin{equation}\label{eq2_5}
    \rang\left\|
    \begin{array}{cccccc}
        pr+\nu_3 & 0 & 0 & 0 & 0 & p(pr+\nu_3)\\
        0 & -1 & 0 & 0 & p & r\\
        r & 2p & \nu_2 &  0 & \nu_1 & \nu_3\\
        \nu_3 & \nu_1 & 0 & \nu_2 & 0 & 0
    \end{array}
    \right\|<4.
\end{equation}
Calculating the determinant generated by the first columns, we obtain $(pr+\nu_3)\nu_2=0$. If $pr+\nu_3=0$, then by virtue of \eqref{eq2_4} the polynomial \eqref{eq1_9} has a multiple root, and the condition \eqref{eq1_14} is satisfied with $x=p$. Let
\begin{equation}\label{eq2_6}
   \nu_2=0, \qquad pr+\nu_3\neq 0.
\end{equation}
Then \eqref{eq2_5} takes the form $(p^2+\nu_1)(2p\nu_3-r\nu_1)=0$. The case $p^2+\nu_1=0$ together with \eqref{eq2_3}, \eqref{eq2_6} gives \eqref{eq1_10}. Supposing
\begin{equation}\label{eq2_7}
   2p\nu_3-r\nu_1=0,
\end{equation}
and using \eqref{eq1_4}, \eqref{eq2_3} and \eqref{eq2_6} we come to $h-2l^2=-(p^2+\nu_1)$,  $k=(p^2+\nu_1)^2$. Then \eqref{eq1_11} holds. Thus all critical values obtained from \eqref{eq2_3} correspond to the Appelrot classes.

Let us investigate another possibility
\begin{equation}\label{eq2_8}
  r=0, \quad \nu_3=0, \quad q \neq 0.
\end{equation}
In the matrix \eqref{eq2_1}, the last two columns are zeros. The remaining fourth-order determinant should equal zero. This yields $2(p^2-q^2+\nu_1)\nu_1\nu_2 - (2pq+\nu_2)(\nu_1^2-\nu_2^2)=0$. Then we can introduce the undetermined multiplier $\varkappa$ putting
\begin{equation}\label{eq2_9}
  p^2-q^2+\nu_1=\varkappa(\nu_1^2-\nu_2^2), \qquad 2pq+\nu_2=2\varkappa\nu_1\nu_2.
\end{equation}
Substituting these expressions into \eqref{eq1_5} and using \eqref{eq1_4} we obtain
\begin{equation}\label{eq2_10}
  \varkappa=\pm\sqrt{k}.
\end{equation}
From \eqref{eq1_4} and \eqref{eq2_9} we have
\begin{equation}\label{eq2_11}
  (p^2+q^2)^2=(\varkappa \nu_1-1)^2+\varkappa^2\nu_2^2.
\end{equation}
Let us rewrite the system \eqref{eq2_9} in the form
\begin{equation*}
  \begin{array}{c}
    (\varkappa \nu_1-1)(p\nu_1+q\nu_2)+\varkappa \nu_2(q\nu_1-p\nu_2)=p(p^2+q^2),\\
    \varkappa \nu_2(p\nu_1+q\nu_2)-(\varkappa \nu_1-1)(q\nu_1-p\nu_2)=q(p^2+q^2).
  \end{array}
\end{equation*}
Then taking into account \eqref{eq2_11} we obtain
\begin{equation*}
  \nu_1=\frac{p^2-q^2+\varkappa}{\varkappa^2-(p^2+q^2)^2}, \quad \nu_2=\frac{2pq}{\varkappa^2-(p^2+q^2)^2}.
\end{equation*}
Substitute the above expressions together with \eqref{eq2_8} into \eqref{eq1_2} -- \eqref{eq1_4} to obtain
\begin{equation}\label{eq2_12}
\begin{array}{l}
\displaystyle h=p^2+q^2-\frac{p^2-q^2+\varkappa}{\varkappa^2-(p^2+q^2)^2}, \quad \displaystyle l=\frac{p}{\varkappa-p^2-q^2},\\
(p^2+q^2)^4-(1+2\varkappa^2)(p^2+q^2)^2-2\varkappa(p^2-q^2)-\varkappa^2(1-\varkappa^2)=0.
\end{array}
\end{equation}
Then $h-2l^2+\varkappa=0$. By virtue of \eqref{eq2_10} the last equation gives \eqref{eq1_11}. Thus, the critical values obtained from the condition \eqref{eq2_8} also correspond to the Appelrot classes.

We now show that the assumption
\begin{equation}\label{eq2_13}
  r^2+\nu_3^3\neq 0, \qquad q \neq 0
\end{equation}
applied to the system \eqref{eq1_2} -- \eqref{eq1_5} gives no new critical values.
First, we note the identity $R(x_1,x_2)=(pr+\nu_3)^2+q^2r^2$ which follows from \eqref{eq1_2} -- \eqref{eq1_5}, \eqref{eq1_7}, \eqref{eq1_8}. Thus, under the condition \eqref{eq2_13} we have
\begin{equation}\label{eq2_14}
  R(x_1,x_2)\neq 0, \qquad x_1 \neq x_2.
\end{equation}
In addition to \eqref{eq1_7}, let us introduce  the variables $\xi_1=\eta_1+i\eta_2$, $\xi_2=\eta_1-i\eta_2$ \cite{bib01}. Due to \eqref{eq2_2} they are connected by non-degenerate transformation with $\nu_1$, $\nu_2$. Eliminating the values $r$, $\nu_1$ in \eqref{eq1_2} -- \eqref{eq1_5}, we obtain two relations \cite{bib01}
\begin{equation}\label{eq2_15}
\begin{array}{c}
  \xi_1 \xi_2=k, \\
  R(x_2)\xi_1+R(x_1)\xi_2+R_1(x_1,x_2)+(x_1-x_2)^2k=0,
\end{array}
\end{equation}
where
\begin{equation}\label{eq2_16}
  R_1(x_1,x_2)=-2hx_1^2x_2^2-4l(x_1+x_2)x_1x_2-(1-k)(x_1+x_2)^2+2(1-k)h-4l^2.
\end{equation}
It follows from the first inequality \eqref{eq2_13} that the condition \eqref{eq2_1} is equivalent to the system determining the critical points \eqref{eq2_15}
\begin{eqnarray}
&  R(x_1)\xi_2=R(x_2)\xi_1, \label{eq2_17}\\
&  \begin{array}{c}
      R'(x_1)\xi_2+\partial R_1(x_1,x_2)/\partial x_1+2(x_1-x_2)k=0, \\
      R'(x_2)\xi_1+\partial R_1(x_1,x_2)/\partial x_2-2(x_1-x_2)k=0.
   \end{array}\label{eq2_18}
\end{eqnarray}
Following \cite{bib01}, we express $\xi_1$, $\xi_2$ from \eqref{eq2_15}
\begin{equation}\label{eq2_19}
    \begin{array}{c}
      2R(x_2)\xi_1=-\left[R_1(x_1, x_2)-(x_1-x_2)^2k\right]+W(x_1,x_2), \\
      2R(x_1)\xi_2=-\left[R_1(x_1, x_2)-(x_1-x_2)^2k\right]-W(x_1,x_2).
     \end{array}
\end{equation}
Here
\begin{equation*}
  W(x_1,x_2)= \left\{ \left[ R_1(x_1, x_2)-(x_1-x_2)^2 k \right]^2-4k R^2(x_1,x_2) \right\}^{1/2}.
\end{equation*}
The condition \eqref{eq2_17} thus yields $W(x_1,x_2)=0$. Using the notation \eqref{eq2_10} we obtain
\begin{equation}\label{eq2_20}
    R_1(x_1,x_2)-(x_1-x_2)^2\varkappa^2=2\varkappa R(x_1,x_2).
\end{equation}
Moreover, substituting $\xi_1$, $\xi_2$ in \eqref{eq2_18} by their expressions obtained from \eqref{eq2_19} we get
\begin{eqnarray}
& &    \begin{array}{l}
        2(1-\varkappa^2)^2+2(1-\varkappa^2)\left[ h^2-1+x_1^2x_2^2+3l(x_1+x_2)\right] \\
        \qquad +2 h^2x_1^2x_2^2+ h\left[-(x_1+x_2)^2+4l(x_1+x_2)x_1x_2-4l^2 \right] \\
        \qquad +(x_1^2+x_2^2)x_1x_2+2l(x_1^2x_2^2-2)(x_1+x_2)+4l^2(x_1^2+3x_1x_2+x_2^2)=0,
     \end{array}\label{eq2_21}\\
& &  \left[R_1(x_1,x_2)-(x_1-x_2)^2\varkappa^2 \right]l+(x_1+x_2+2lx_1x_2)R(x_1,x_2)=0.\label{eq2_22}
\end{eqnarray}
From \eqref{eq2_20}, \eqref{eq2_22} we have $\left[x_1+x_2+2l(x_1x_2+\varkappa)\right]R(x_1,x_2)=0$. Therefore the assumption \eqref{eq2_14} yields
\begin{equation}\label{eq2_23}
  x_1+x_2=-2l(x_1x_2+\varkappa).
\end{equation}
Let us substitute the expression obtained for $x_1+x_2$ into \eqref{eq2_21}, \eqref{eq2_22}. Then
\begin{equation*}
  \begin{array}{l}
    (\varkappa-h+2l^2)\left[ 4l^2(x_1x_2+\varkappa)^2-(h+\varkappa)x_1^2x_2^2+1-\varkappa^2)\right]=0, \\
    (\varkappa-h+2l^2)\left[(x_1x_2+\varkappa)^2-1 \right]=0.
  \end{array}
\end{equation*}
If $\varkappa-h+2l^2=0$, then according to \eqref{eq2_10} we again come to \eqref{eq1_13}. Another possibility is $x_1x_2=\pm1-\varkappa$, $x_1+x_2=\mp2l$, and $2l^2=(h+\varkappa)(1\mp\varkappa)$. Then the direct check gives $R(x_1, x_2)=0$; this case is excluded by \eqref{eq2_14}.

Finally, we proved that the bifurcation set $\Sigma$ is the part of the discriminant surface of the polynomial $\Phi(s)$ corresponding to real solutions of \eqref{eq1_2} -- \eqref{eq1_5}.

Note that the set of critical points consists of those trajectories of \eqref{eq1_1} which, in Appelrot's terminology, correspond to \textit{especially remarkable motions} (one of the variable \eqref{eq1_6} remains constant during the motion).

\mysec{3}
The form of equations \eqref{eq1_11}, \eqref{eq1_14} implies that it is convenient to consider the cross sections $\Sigma_l\subset \bbR^2(k,h)$ of the set $\Sigma$ by the planes $l=\cons$. This method is more convenient for the aim of the present investigation than the method used in \cite{bib03}, where, for instance, the solutions of \eqref{eq1_12} are considered relative to $l$ and the projections of the discriminant surface on the planes $lh$ and $lk$ are studied. Since $\Sigma_l=\Sigma_{-l}$, we restrict ourselves to the case $l\geqslant 0$.

Equation \eqref{eq1_11} in the $kh$-plane defines the parabola with the vertex at $(0, 2l^2)$.

Consider the curve \eqref{eq1_12}. When $l=0$ it splits into the straight line $k=1$ and the parabola $k=1+h^2$. For $l>0$ let us use \eqref{eq1_14}. The curve investigated has a cusp at $x=(-l/2)^{1/3}$ and a vertical asymptote $k=1$ $(x\rightarrow\pm 0$, $h\rightarrow\mp\infty)$. As $x\rightarrow\pm\infty$, both coordinates $k$ and $h$ approach $+\infty$ so that $(k(x),h(x))$ asymptotically approaches the correspondent curve $k=h^2\pm 4l\sqrt{h}+1$.

In the general case the curve \eqref{eq1_14} has two points of intersection with the parabola \eqref{eq1_11}
\begin{eqnarray}
& &  (k,h)=\left( (l^2+1)^2, l^2-1 \right), \qquad x=l, \label{eq3_1}\\
& &  (k,h)=\left( (l^2-1)^2, l^2+1 \right), \qquad x=l \label{eq3_2}
\end{eqnarray}
and a tangency point
\begin{equation}\label{eq3_3}
  (k,h)=\left(\frac{1}{16l^4},\frac{1}{4l^2}+2l^2\right), \qquad  x=-\frac{1}{2l}.
\end{equation}
The exception is the case $l^2=1/2$ when the points \eqref{eq3_2} and \eqref{eq3_3} coincide with the cusp. For $l^2=4/\left(3\sqrt{3}\right)$ the cusp is on the axis $k=0$ (which is obviously a part of the boundary of the region of possible motions). One more singularity arises for $l=1$ when the point \eqref{eq3_2} passes from one branch of the parabola \eqref{eq1_11} to another.

It is interesting to observe the transformation of the curve \eqref{eq1_14} as $l\rightarrow 0$. The branch $-\infty < x < 0$ converts to the half-line $\{k=1, h\geqslant 0\}$ and the upper part of the parabola $k=h^2+1$ $(h\geqslant 0)$. Consequently, the cusp reaches the point $(1,0)$. The branch $0<x<+\infty$ joins with the half-line $\{k=1, h\leqslant 0\}$ and the same upper part of the parabola $k=h^2+1$. The tangency point of the curve \eqref{eq1_14} and the parabola \eqref{eq1_11} moves to infinity as $l\rightarrow 0$.

The form of the sets $\Sigma_l$ is shown in Fig. 1 -- 4 for the following values of the area constant: $0<l^2<1/2$; \ $1/2<l^2<4/(3\sqrt{3})$; \ $4/(3\sqrt{3})<l^2<1$; \ $l^2>1$.

\begin{table}
\centering
{\small
\begin{tabular}{cc}
\includegraphics[width=0.3\textwidth]{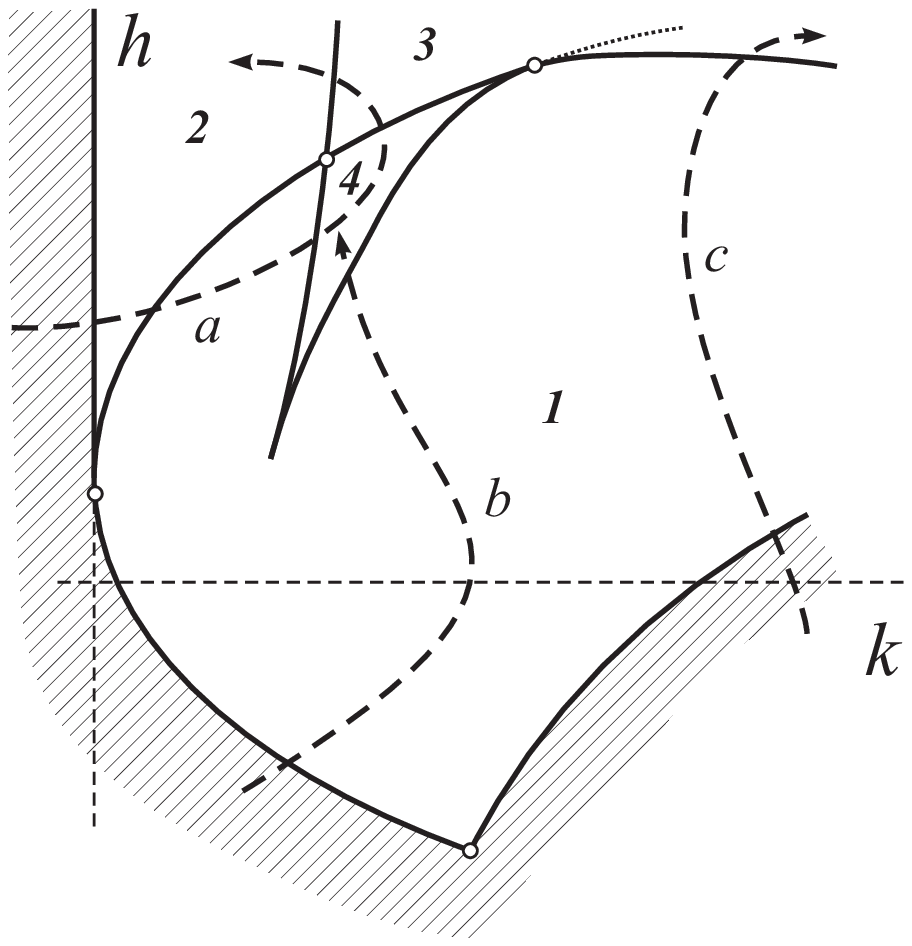} & \includegraphics[width=0.3\textwidth]{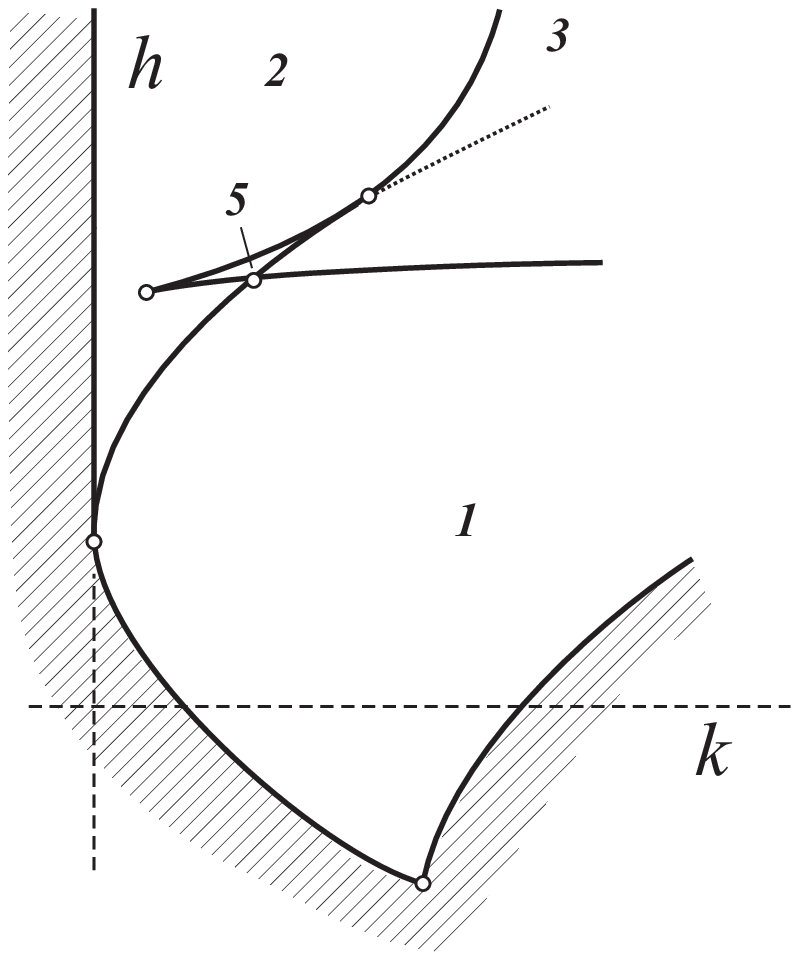}\\
\textit{Fig}.~1 & \textit{Fig}.~2 \\
\includegraphics[width=0.3\textwidth]{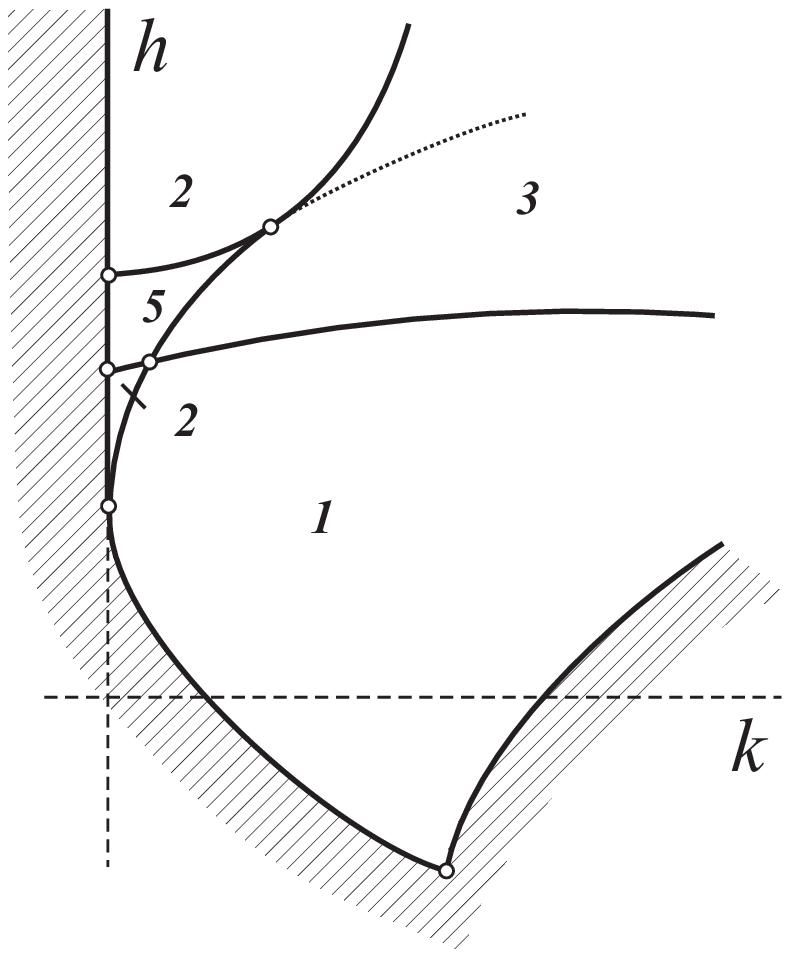} & \includegraphics[width=0.3\textwidth]{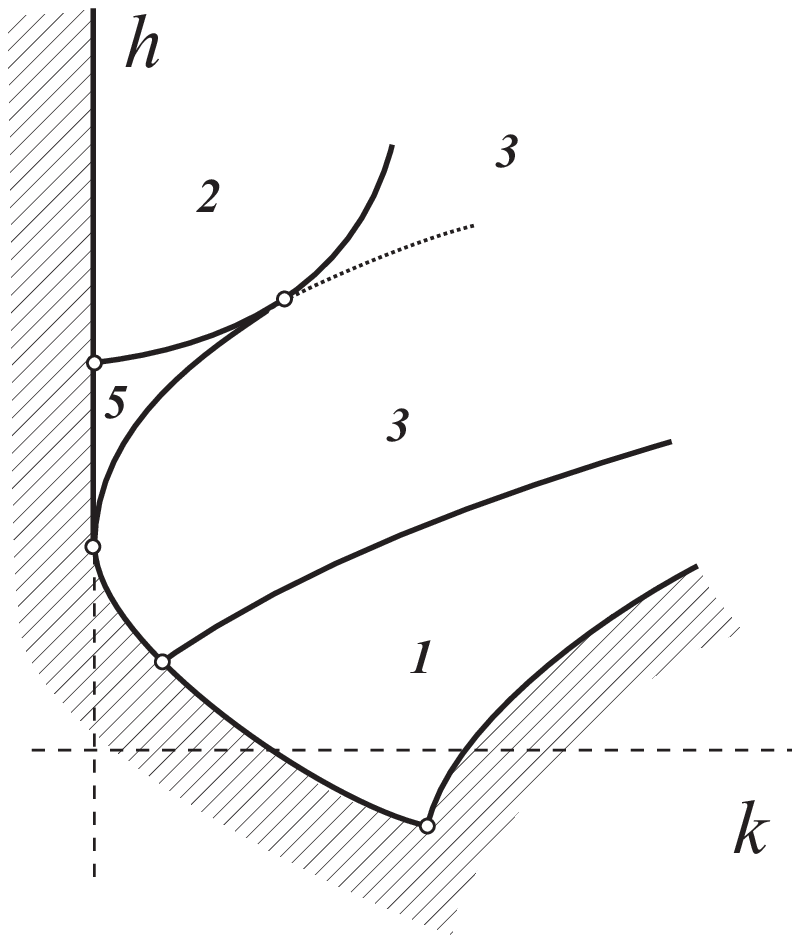}\\
\textit{Fig}.~3 & \textit{Fig}.~4
\end{tabular}
}
\end{table}

\mysec{4} As it was already mentioned above, the topological type of the manifold $J_{k,h,l}$ can change only when the point $(k,h,l)$ passes through the critical value, i.e., through the set \eqref{eq1_10} -- \eqref{eq1_12}. Consequently, in the shaded regions of the $kh$-plane containing points with $k<0$ or $h<-1$ we have $J_{k,h,l}=\varnothing$.

We will show that $\Sigma_l$ does not include the part of the upper branch
\begin{equation}\label{eq4_1}
  h=2l^2+\sqrt{k}
\end{equation}
of the parabola \eqref{eq1_11} lying to the right of the tangency point \eqref{eq3_3} of the parabola and the curve \eqref{eq1_14}. It is sufficient to show that under the condition \eqref{eq4_1} real critical points exist only for the values
\begin{equation}\label{eq4_2}
  \sqrt{k}\leqslant 1/(4l^2).
\end{equation}
This fact corresponds to the assertion of \cite{bib02, bib03} stating that the third class does not contain especially remarkable motions when $\sqrt{k}> 1/(4l^2)$.
As we have seen above, the values \eqref{eq4_1} are critical in the following three cases. The first one takes place if  \eqref{eq2_3}, \eqref{eq2_6} and \eqref{eq2_7} are satisfied
\begin{equation}\label{eq4_3}
  2p\nu_3-r\nu_1=0, \qquad q=0, \qquad  \nu_2=0.
\end{equation}
The second case corresponds to the conditions \eqref{eq2_12} with $\varkappa=-\sqrt{k}$. The third one is the case of \eqref{eq2_23} with $\varkappa=\sqrt{k}$. The last two cases lead to the equation
\begin{equation}\label{eq4_4}
  l(p^2+q^2)+p+l\sqrt{k}=0,
\end{equation}
which has real solutions for $p$ and $q$ if and only if the equation $lp^2+p+l\sqrt{k}=0$ has real roots in $p$. This leads to the inequality \eqref{eq4_2}.

Let us consider the case \eqref{eq4_3}. From \eqref{eq1_3} and the first equation of \eqref{eq4_3} we have $\nu_1=4lp/(4p^2+q^2)$, $\nu_3=2lp/(4p^2+q^2)$. Substitution into \eqref{eq1_4} gives $4p^2+q^2=4l^2$. Then \eqref{eq1_5} takes the form \eqref{eq4_4}.

In Fig. 1 -- 4, the regions cut out from the same connected component of $\bbR^3 \setminus \Sigma$ are denoted by the same numbers. Thus, we have five components in which integral manifolds are nonempty. To establish the number of tori in $J_{k,h,l}$, let us consider the image of $J_{k,h,l}$ in the plane $pq$. In \cite{bib02}, this image is called the region of real motions. It is not difficult to establish the connection between the regions 1 -- 5 in $\bbR^3 \setminus \Sigma$ and the cases considered in \cite{bib02}. As a result, for the regions 1 -- 5, we obtain the projections on the $pq$-plane shown in Fig. 5,\,\textit{а} -- \textit{e} respectively. Here the $q$ axis is vertical, and $x_i$ are the real roots of the polynomial $R(x)$.

\begin{table}
\centering
{\small
\begin{tabular}{c}
\includegraphics[width=0.4\textwidth]{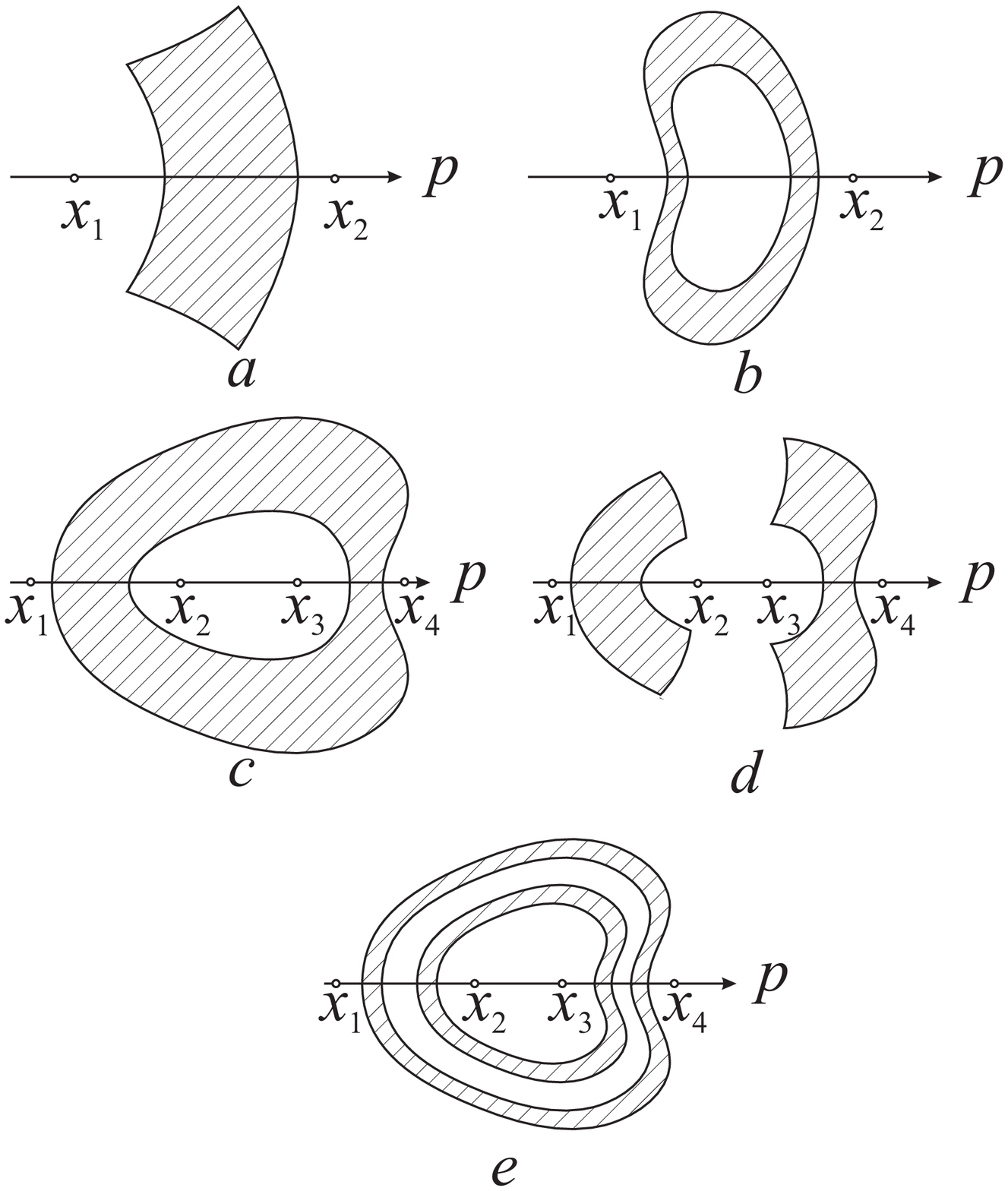} \\
\textit{Fig}.~5
\end{tabular}
}
\end{table}

Let us find the number of the pre-images of the points in the shaded sets. Since all of them reaches the $p$ axis, it is sufficient to do this for $q=0$. From \eqref{eq1_2} -- \eqref{eq1_5} using the notation \eqref{eq1_8} and \eqref{eq2_16} we have
\begin{equation}\label{eq4_5}
    \begin{array}{ll}
        \nu_1=p^2+r^2/2-h,  & r\nu_3=-pr^2+2(l+hp-p^3), \\
        4R(p)r^2=[R'(p)]^2, &  4R^2(p)\nu_2^2=4kR^2(p)-R_1^2(p,p).
    \end{array}
\end{equation}
Take the segment of intersection of the region of real motions with the $p$-axis.
At the inner point of it $q=0,p=p_0,R(p_0)\ne 0,R'(p_0)\ne 0$, and the system \eqref{eq4_5} has four solutions with respect to $(r, \nu_1, \nu_2, \nu_3)$. Thus to every inner point of the regions shown in Fig. 5, four points of the integral manifold are projected. Hence the region diffeomorphic to a ring is covered by two tori, and the region diffeomorphic to a rectangle is covered by one torus. Finally, in component 1 of the set $\bbR^3 \setminus \Sigma$ the integral manifold consists of one two-dimensional torus, in components 2 -- 4 it consists of two, and in component~5 it consists of four two-dimensional tori.

To establish the character of bifurcations, we analyze the regions of real motions at the points of $\Sigma$ (the boundaries of these regions are given in \cite{bib03}). We find out the possible changes of the number of the pre-images of the points in these regions with respect to the projection of the integral surfaces on the $pq$-plane. The integral surfaces in this case are not smooth manifolds. The details of the method are given in \cite{bib12}.

Let us introduce the following notation: $S$ is a set homeomorphic to the circle; $V = S \bigvee S$ is the figure eight curve; $W$ is a set homeomorphic to the intersection of a two-dimensional sphere with a pair of planes passing through its centre (two circles transversally intersecting at two points); $P$ is the non-trivial fiber bundle over the circle with the fiber equal to the figure eight curve. The surface $P$ is obtained from $[0; 1] {\times} V$ by the identification $\{0\} {\times}V$ with $\{1\} {\times} V$ with respect to a mapping that is homotopic to the central symmetry of the eight figure curve. Denote $Q=W{\times} S$ and $U=V {\times} S$.

Let $\gamma$ be the boundary of the $\varepsilon$-neighbourhood of the surface $P$ embedded in $\bbR^3$. Then obviously $\gamma = 2T^2$. It is easy to imagine a one-parameter set of surfaces $P_\tau$, $\tau\in [-\varepsilon,\varepsilon]$ such that $P_\tau = T^2$ when $\tau\neq 0$ and $P_0=P$. The type of the bifurcation $T^2\rightarrow P \rightarrow T^2$ as $\tau$ changes in $[-\varepsilon,\varepsilon]$ will be denoted by $(1, 1)$. Similarly we define the bifurcation of the types
\begin{equation*}
  \begin{array}{ll}
     (2, 2): & 2T^2\rightarrow Q \rightarrow 2T^2; \\
     (1, 2): & T^2\rightarrow U \rightarrow 2T^2;\\
    (0, 1): & \varnothing \rightarrow S \rightarrow T^2.
   \end{array}
\end{equation*}
The last two bifurcations occurring in the reverse order we denote, respectively, by the symbols $(2, 1)$ and $(1, 0)$. The notation $(1{:}1)$ stands for a continuous deformation of the connected component of the integral manifold on which there are no critical points. The symbols of simultaneously occurring bifurcations are connected by the plus sign, or indicated with an integral multiplier, if they are identical.

We now give a list of the bifurcation sequences taking place along the dashed arrows in Fig.~1: a)~$2(0, 1)$, $(2, 1)$, $(1, 2)$, $(2, 2)$, $2(1, 1)$; b)~$(0, 1)$, $(1{:}1) + (0, 1)$; c)~$(0, 1)$, $(1, 2)$. The transition from component 2 to component 5 from above (Fig.~3) is accompanied by the bifurcation $2(1{:}1) + 2(0, 1)$, and from below by $2(1, 2)$. Passing from component 5 to component 3 we have the bifurcation $2(2, 1)$, and the exit from component 5 into the region  ${k < 0}$ gives the bifurcation $4(1, 0)$.

\end{document}